\documentclass{article}

\usepackage{microtype}
\usepackage{graphicx}
\usepackage{subfigure}
\usepackage{booktabs} 
\graphicspath{ {./} }

\usepackage{hyperref}
\usepackage{multirow}
\usepackage{array}

\usepackage[accepted]{icml2021}

\icmltitlerunning{APObind: A Dataset of Ligand Unbound Protein Conformations for Machine Learning Applications in De Novo Drug Design}

\begin{document}

\twocolumn[
\icmltitle{APObind: A Dataset of Ligand Unbound Protein Conformations for Machine Learning Applications in De Novo Drug Design}




\begin{icmlauthorlist}
\icmlauthor{Rishal Aggarwal}{iiit}
\icmlauthor{Akash Gupta}{iiit}
\icmlauthor{U. Deva Priyakumar}{iiit}
\end{icmlauthorlist}

\icmlaffiliation{iiit}{International Institute of Information Technology, Hyderabad 500 032, India}

\icmlcorrespondingauthor{Rishal Aggarwal}{rishal.aggarwal@research.iiit.ac.in}

\icmlkeywords{Machine Learning, ICML}

\vskip 0.3in
]



\printAffiliationsAndNotice{} 

\begin{abstract}
Protein-ligand complex structures have been utilised to design benchmark machine learning methods that perform important tasks related to drug design such as receptor binding site detection, small molecule docking and binding affinity prediction. However, these methods are usually trained on only ligand bound (or holo) conformations of the protein and therefore are not guaranteed to perform well when the protein structure is in its native unbound conformation (or apo), which is usually the conformation available for a newly identified receptor.  A primary reason for this is that the local structure of the binding site usually changes upon ligand binding. To facilitate solutions for this problem, we propose a dataset called APObind that aims to provide apo conformations of proteins present in the PDBbind dataset, a popular dataset used in drug design. Furthermore, we explore the performance of methods specific to three use cases on this dataset, through which, the importance of validating them on the APObind dataset is demonstrated.
\end{abstract}

\section{Introduction}

The structure based drug design (SBDD) paradigm involves the analysis of protein structures for the rational design of drug molecules that can form stable complexes with it \cite{anderson2003process}. SBDD follows multiple steps including the identification of druggable and functional binding sites on the receptor surface, screening large libraries for candidate lead molecules and de novo design of ligand molecules.

Data-driven machine learning (ML) and deep learning (DL) models have shown state-of-the-art performance in general data domains such as computer vision \cite{7780459} and natural language processing (NLP) \cite{Lan2020ALBERT:} leading to their increasing adoption in developing benchmark methods for several chemical and biological tasks like drug design \cite{Vamathevan2019}. A drawback of these methods however is that, they tend not to generalise well to data that does not resemble the data distribution used for training. The viability of such models therefore depend on well curated training data that translates well into real world applications.

Deep Learning models pertaining to SBDD workflows are usually trained on datasets containing 3D structures of protein-ligand complexes \cite{batool2019structure}. PDBbind \cite{doi:10.1021/jm048957q} is a predominantly used dataset that provides experimental binding affinity values for protein-ligand co-crystal structures present in the Protein Data Bank (PDB) \cite{10.1093/nar/28.1.235}. 
Deep learning architectures usually use voxelized \cite{doi:10.1021/acs.jcim.7b00650} or graph like representations \cite{10.1371/journal.pone.0249404} of the 3D structures present in PDBbind for computation to get benchmark performances. However, it is well known that the local structure of the binding site changes upon ligand binding, a phenomenon usually referred to as "induced fit" \cite{koshland1995key}. Therefore, since PDBbind contains only the ligand bound conformation (holo) of the protein structures, these methods are not expected to perform well when applied to proteins in their native unbound conformations (apo). 

In this work we propose a new dataset called APObind that provides apo structures for proteins present in the PDBbind database. To the best of our knowledge, APObind forms the largest collection of apo conformations with holo counterparts along with binding affinity values. The dataset is designed to facilitate robust validation of methods that are intended to work on apo structures, such as binding site detection and affinity prediction. Moreover, we show that the methods that work on holo structures don't translate well to apo structures by showing three use cases - ligand docking, protein-ligand scoring and protein binding site detection. By keeping this diversity in problem space, we not only demonstrate the importance of APObind but also establish a generalized claim on the importance of inclusion of the apo conformations when training deep learning models without being biased towards a specific task.  


\begin{figure*}[t]
\begin{center}

\includegraphics[width=0.8\textwidth]{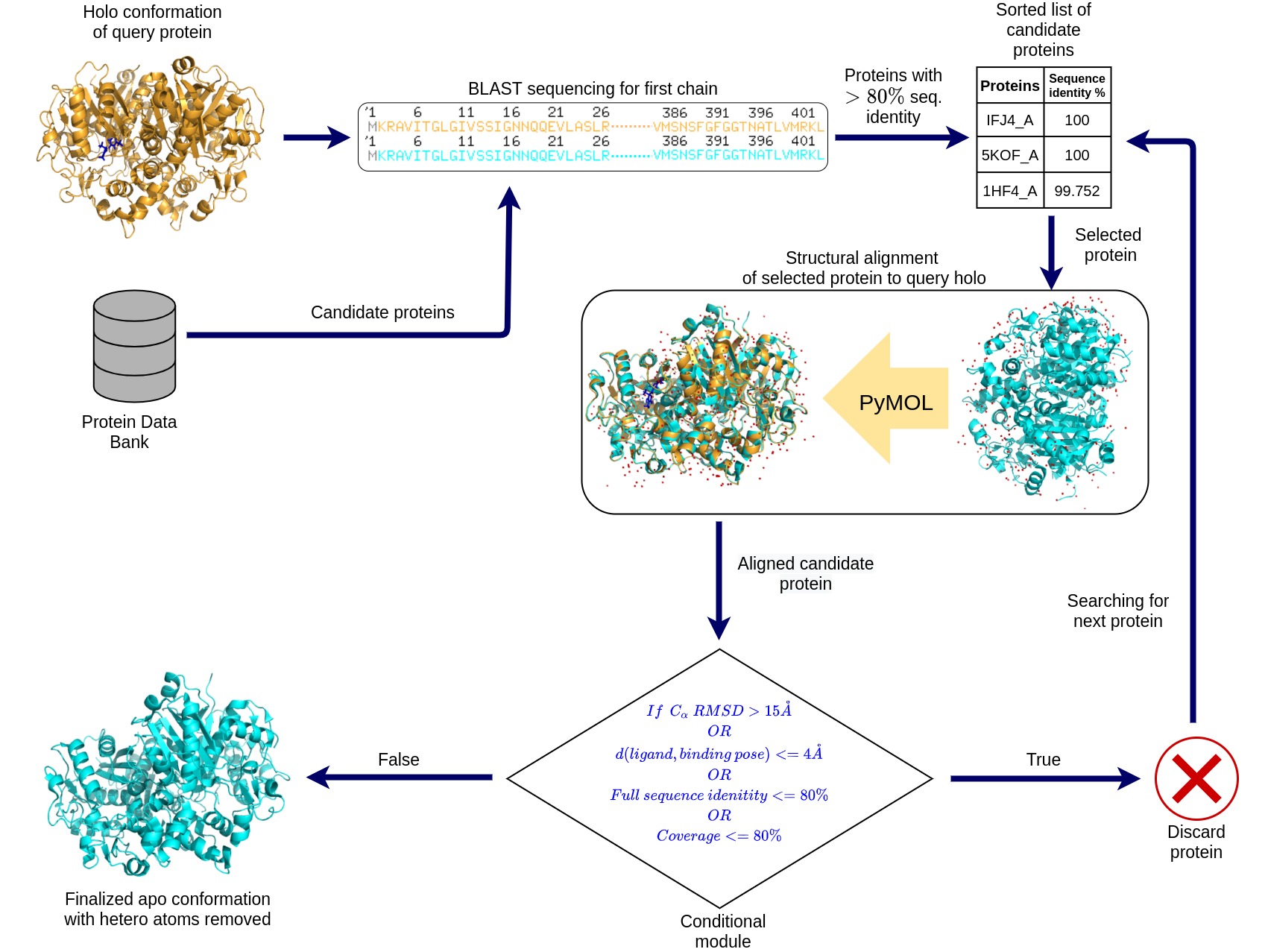}
\caption{An illustration of the pipeline used for preparing APObind. Holo proteins are shown in "bright orange", apo proteins in "cyan", ligand in "blue", hetero atoms in "red".}
\label{fig:datapipeline}
\end{center}
\end{figure*}

\section{Methods}
\label{sec:methods}
\subsection{Data and Preparation}
\label{sec:methods1}

Figure 1 depicts the pipeline for obtaining apo conformations of protein structures. Candidate apo structures for holo equivalents are initially prepared by querying the sequence of the first chain of the holo structure through PDB using the BLAST program \cite{10.1093/nar/25.17.3389} and retaining all the hits that displayed greater than 80\% sequence identity. These proteins are then sorted in order of identity and each structure is checked for a series of conditions until either a suitable structure is identified or all the BLAST hits are rejected. Each hit protein is structurally aligned to the holo structure via superposition of corresponding C$_{\alpha}$ atom of amino acid residues using PyMOL \cite{delano2002pymol}. Post alignment, if the backbone C$_{\alpha}$ Root Mean Square Deviations (RMSD) of aligned residues is greater than 15 {\AA}, or if the full protein sequence alignment showed lesser than 80\% sequence identity or sequence coverage, then the structure is rejected. Furthermore, if any of the ligands of the hit structure are within 4 {\AA} of any atoms of the crystal structure pose of the complex then the structure is rejected. Finally, if a hit protein passes all these conditions, then all hetero atoms of that structure are removed and only the chains involved in alignment are saved. This is then labelled as the apo equivalent of the holo structure.  

Since we mainly intend to work with small molecule binding sites, any protein-ligand complex with ligand molecule having a molecular weight greater than 1000 Daltons was removed from this analysis. Using this procedure, we obtained ligand-free protein structure equivalents for 10,599 out of 16,608 protein-ligand complexes in the PDBbind database. Test and train data splits for deep learning models were made by clustering proteins based on 70\% sequence identity. This resulted in a training set of 7638 data points and test set of 2961 data points.

\begin{figure*}[t]
\centering
\subfigure[]{\includegraphics[width=.4\textwidth]{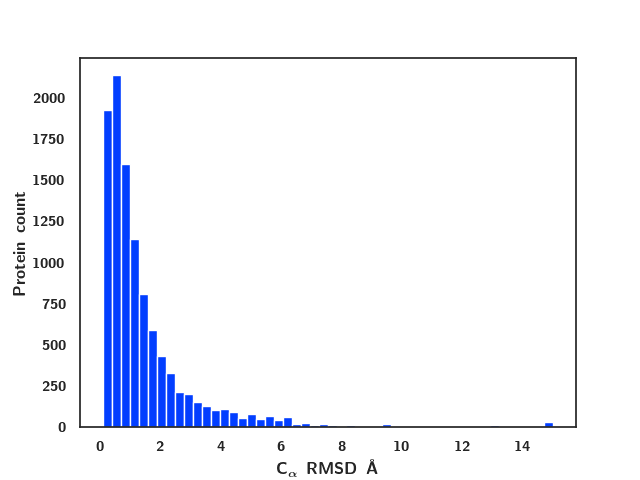}}
\subfigure[]{\includegraphics[width=.4\textwidth]{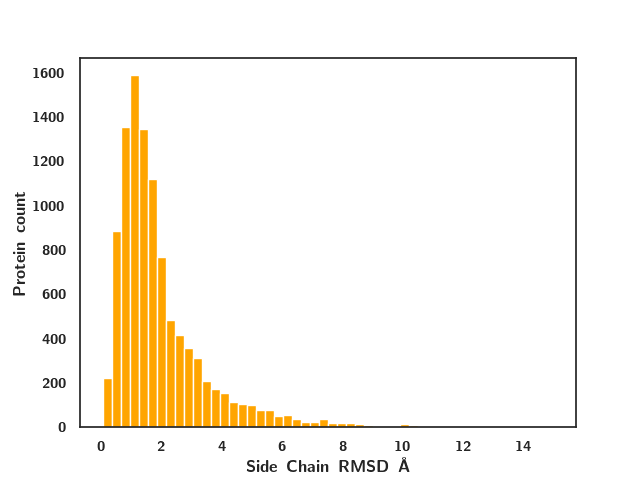}}
\caption{Distribution of heavy atom RMSD between apo and holo structures for (a) C$_\alpha$ Backbone and (b) Side Chain groups in the binding site}
\label{fig:RMSD}
\end{figure*}

\subsection{Implementation of Methods}
\label{sec:methods2}

Docking is performed using the smina \cite{koes2013lessons} software which uses the Autodock Vina \cite{Trott2010} scoring function and provides a convenient command line interface (CLI) for small molecule docking. Ligands are redocked on both the apo and holo structures in the same locality as the crystal pose structure. 

Pafnucy \cite{10.1093/bioinformatics/bty374} is implemented to evaluate the performance of a protein-ligand scoring function on APObind. It is a convolutional neural network that takes voxelized input of the protein-ligand complex as input and predicts the binding affinity. The ligand is placed in the same binding position as in the holo structure for the apo equivalents. Three versions of the dataset are prepared for model training, (1) APO-only, containing only apo conformations from APObind; (2) HOLO-only, containing only holo conformations from PDBbind; and (3) BOTH, containing apo and holo conformations from both the datasets.

DeepPocket \cite{aggarwal_gupta_chelur_jawahar_priyakumar_2021} is used for predicting binding sites in 3D protein structures. DeepPocket uses a geometry based binding site detection software known as Fpocket \cite{le2009fpocket} to parse the protein structure and return candidate binding pocket centers. These candidate centers are then ranked using a classification model that is trained to identify ligand binding sites. Similar to Pafnuncy, three models of DeepPocket are trained on APO-only, HOLO-only and BOTH versions of the dataset. DeepPocket performance is measured using the DCA criterion that reports the percentage of times top ranked pocket centers are within 4 {\AA} of any ligand heavy atom. 


\section{Results}


\subsection{Structural Differences Between Apo and Holo}

The significance of conformations in various structure-based prediction tasks is highlighted when analyzing the extent of differences in the structural changes in a protein in the process of ligand binding. The protein backbone forms the overall 3D structure of the protein and is directly related to the manner in which all of its domains are arranged. Side chains of residues present in the binding site on the other hand are responsible for interactions with the ligand. Changes in the orientation as a result of a binding process in any of these components may result in major structural changes of the protein. To examine this, we plot the distribution of the heavy atom RMSD of (a) the complete protein C$_{\alpha}$ backbone and (b) the binding pocket side chains, between holo and their corresponding apo proteins in Figure \ref{fig:RMSD}. Binding site residues here are defined as any residue within 6 {\AA} of any ligand heavy atom. A majority of the structures are within 8 {\AA} RMSD for both the distributions indicating similar structures with conformational changes have been found through the apo search. 

\begin{figure}[!h]

\subfigure[]{\includegraphics[width=.23\textwidth]{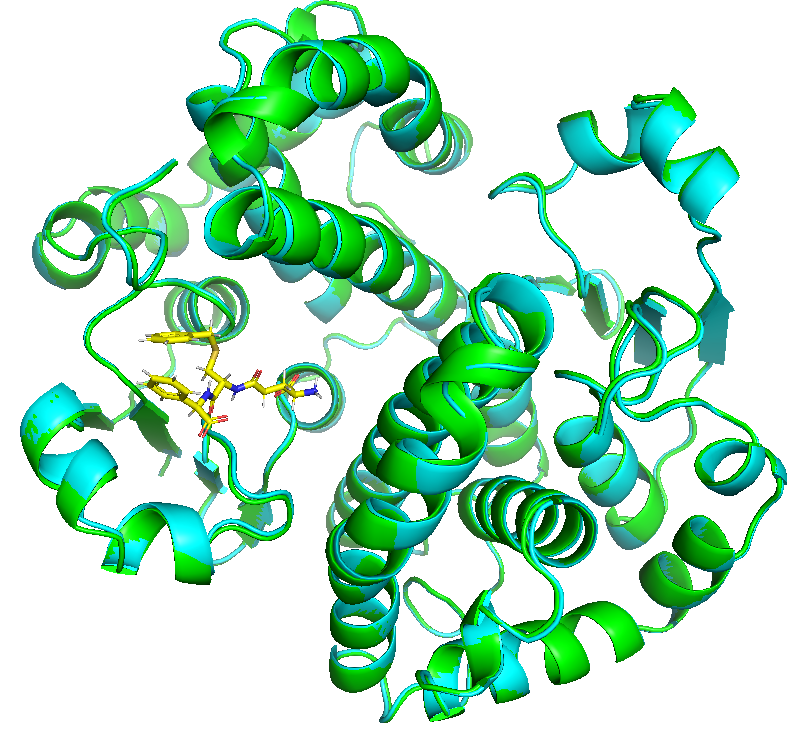}}
\subfigure[]{\includegraphics[width=.23\textwidth]{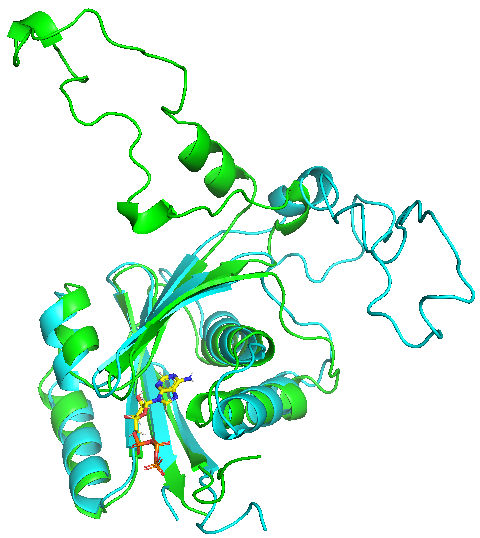}}

\caption{Aligned holo and apo structures of (a) Human Glutathione S-Transferase protein (PDB ID: 10GS) and (b) Human Menkes protein  (PDB ID: 2KMX). Apo structures are shown in "cyan" and holo structures in "green" with the bound ligand in "yellow" }
\label{fig:aligned prots}
\end{figure}

For a more qualitative assessment of the retrieved structures, we also show visualizations of apo and holo protein structures corresponding to high and low C$_\alpha$ backbone RMSD in Figure \ref{fig:aligned prots}. We note that the Human Menkes protein  (PDB ID: 2KMX) undergoes a large structural change upon ligand binding resulting in a C$_\alpha$ backbone RMSD value of 12.29 {\AA}, as opposed to the protein human Glutathione S-Transferase (PDB ID: 10GS) which remains largely the same with C$_\alpha$ backbone RMSD = 0.28 {\AA}.




\subsection{Ligand Docking}

Table \ref{smina} shows the smina performance on both apo proteins from APObind and their holo counterparts separately. It reports the percentage of times smina generated a pose within 2.5 {\AA} of the ligand binding pose in the Top-n ranks among all the generated poses. The BEST metric takes all generated poses into consideration.   

\begin{table}[h]
\caption{Percentages of times a docked pose with a RMSD within 2.5 {\AA} of the crystal structure pose is generated in Top-n ranks.}
\label{smina}
\begin{center}
\begin{small}
\begin{sc}
\begin{tabular}{ccccc}
\toprule
{} & Top-1 & Top-3 & Top-5 & Best \\
\midrule
APO & 13.74\% & 22.55\% & 27.48\% & 37.73\% \\
HOLO & 57.94\% & 71.49\% & 76.49\% & 84.73\%\\
\bottomrule
\end{tabular}
\end{sc}
\end{small}
\end{center}
\end{table}

We note a huge difference of 44.2\% in the Top-1 scores indicating the inferior performance of Smina on docking ligands on the apo conformations. 
This difference increases even further in Top-3 and Top-5 scores with holo performance going till 76.49\% while apo being limited to 27.48\%. 
The holo performance peaks at 84.73\% while apo stays at 37.73\% which indicates that smina works on most of the proteins in holo conformations but isn't able to find the correct ligand pose for their apo counterparts. This is mainly due to the deviation of side chain groups from their holo binding positions. This highlights the importance of such a dataset as it can be used assess the generalizability of docking methods on non binding conformation during development, especially since most drug design tasks are initiated based on apo structures.

\subsection{Binding Affinity Prediction}



\begin{table}[h]
\caption{Performance metrics of pafnucy on apo and holo proteins in APObind test set}
\vspace{5pt}
\label{pafnucy}
\begin{center}
\begin{small}
\begin{sc}
\begin{tabular}{cccc}
\toprule
Model & Test set & RMSD ({\AA}) & R \\
\midrule
\multirow{2}{*}{APO-only} & APO & 1.601  & 0.485 \\
                          & HOLO & 1.590 & 0.525 \\
\midrule
\multirow{2}{*}{HOLO-only} & APO & 1.667 & 0.452 \\
                          & HOLO & 1.545 & 0.555 \\
\midrule
\multirow{2}{*}{BOTH} & APO & 1.592 & 0.498 \\
                          & HOLO & 1.535 & 0.551 \\
\bottomrule
\end{tabular}
\end{sc}
\end{small}
\end{center}
\end{table}
Table \ref{pafnucy} shows the RMSD and pearson correlation (R) values for the 3 Pafnucy models (described in section \ref{sec:methods2}) for apo and holo conformations on comparing predicted and experimental affinity values. 
As expected, there is a difference in performance (0.122 in RMSD) for the HOLO-only model on the apo and the holo test sets. The model trained on the APO-only dataset shows similar performance on both the holo and apo test sets. On comparing BOTH model with HOLO-only, we found that the performance increases by 0.075 RMSD on apo conformations of the APObind test set and slightly on holo counterparts as well. 
The marginal improvement on adding apo information to HOLO-only model training for both apo and holo test sets indicates the advantage of using such a dataset for binding affinity prediction. We therefore conclude that augmenting the datasets with apo conformations while training boosts the performance at test time on apo as well as on holo proteins.  

\subsection{Binding Site Detection}

Fpocket detected pocket centers that passed the DCA criterion for 95.45\% of holo structures and 86.77\% apo structures. Table \ref{deeppocket} reports the percentage of times the classification models ranked the correct pocket center within the Top-1 and Top-3 ranks along with the AUC-ROC for each test set

\begin{table}[h]
\caption{Performance metrics of DeepPocket on apo and holo proteins in APObind test set}
\label{deeppocket}
\vspace{5pt}
\begin{small}
\begin{sc}
\resizebox{\columnwidth}{!}{%
\begin{tabular}{ccccc}
\toprule
Model & Test set & Top-1 & Top-3 & AUC-ROC\\
\midrule
\multirow{2}{*}{APO-only} & APO & 42.15\% & 61.46\% & 0.8489 \\
                          & HOLO & 53.46\% & 75.46\% & 0.8782\\
\midrule
\multirow{2}{*}{HOLO-only} & APO & 33.54\% & 54.72\% & 0.7919\\
                          & HOLO & 55.76\% & 76.58\% & 0.8962\\
\midrule
\multirow{2}{*}{BOTH} & APO & 38.98\% & 61.39\% & 0.8479 \\
                          & HOLO & 61.39\% & 76.99\% & 0.8931\\
\bottomrule
\end{tabular}
}
\end{sc}
\end{small}
\end{table}

The results show a similar pattern as in binding affinity prediction with the model trained on BOTH clearly outperforming the model trained on HOLO-only for both the conformations in the test set. As before, the model trained on APO-only shows the best performance on the apo test set but the worst performance on the holo set indicating that the addition of the holo set improves the models generalization to holo structures albeit at a slight performance cost on apo structures.

\section{Conclusion}

We have designed APObind, a dataset of apo conformations to represent protein structures used in the initial stages of drug design. The dataset can be used to validate current data driven and machine learning methods on unbound conformations. In addition, the dataset can be used to improve upon current performances therefore leading to models that are more viable to use in drug design applications. Finally, APObind will also be useful for robust validation of methods that are specifically designed to work on apo conformation of the receptor target. 

\section*{Acknowledgements}

We thank IHub-Data, IIIT Hyderabad for financial support.

\bibliography{main}

\begin{thebibliography}{17}
\providecommand{\natexlab}[1]{#1}
\providecommand{\url}[1]{\texttt{#1}}
\expandafter\ifx\csname urlstyle\endcsname\relax
  \providecommand{\doi}[1]{doi: #1}\else
  \providecommand{\doi}{doi: \begingroup \urlstyle{rm}\Url}\fi

\bibitem[Aggarwal et~al.(2021)Aggarwal, Gupta, Chelur, Jawahar, and
  Priyakumar]{aggarwal_gupta_chelur_jawahar_priyakumar_2021}
Aggarwal, R., Gupta, A., Chelur, V., Jawahar, C.~V., and Priyakumar, U.~D.
\newblock Deeppocket: Ligand binding site detection and segmentation using 3d
  convolutional neural networks, May 2021.
\newblock URL
  \url{https://chemrxiv.org/articles/preprint/DeepPocket_Ligand_Binding_Site_Detection_and_Segmentation_using_3D_Convolutional_Neural_Networks/14611146/1}.

\bibitem[Altschul et~al.(1997)Altschul, Madden, Schäffer, Zhang, Zhang,
  Miller, and Lipman]{10.1093/nar/25.17.3389}
Altschul, S.~F., Madden, T.~L., Schäffer, A.~A., Zhang, J., Zhang, Z., Miller,
  W., and Lipman, D.~J.
\newblock {Gapped BLAST and PSI-BLAST: a new generation of protein database
  search programs}.
\newblock \emph{Nucleic Acids Research}, 25\penalty0 (17):\penalty0 3389--3402,
  09 1997.
\newblock ISSN 0305-1048.
\newblock \doi{10.1093/nar/25.17.3389}.
\newblock URL \url{https://doi.org/10.1093/nar/25.17.3389}.

\bibitem[Anderson(2003)]{anderson2003process}
Anderson, A.~C.
\newblock The process of structure-based drug design.
\newblock \emph{Chemistry \& biology}, 10\penalty0 (9):\penalty0 787--797,
  2003.

\bibitem[Batool et~al.(2019)Batool, Ahmad, and Choi]{batool2019structure}
Batool, M., Ahmad, B., and Choi, S.
\newblock A structure-based drug discovery paradigm.
\newblock \emph{International journal of molecular sciences}, 20\penalty0
  (11):\penalty0 2783, 2019.

\bibitem[Berman et~al.(2000)Berman, Westbrook, Feng, Gilliland, Bhat, Weissig,
  Shindyalov, and Bourne]{10.1093/nar/28.1.235}
Berman, H.~M., Westbrook, J., Feng, Z., Gilliland, G., Bhat, T.~N., Weissig,
  H., Shindyalov, I.~N., and Bourne, P.~E.
\newblock {The Protein Data Bank}.
\newblock \emph{Nucleic Acids Research}, 28\penalty0 (1):\penalty0 235--242, 01
  2000.
\newblock ISSN 0305-1048.
\newblock \doi{10.1093/nar/28.1.235}.
\newblock URL \url{https://doi.org/10.1093/nar/28.1.235}.

\bibitem[DeLano et~al.(2002)]{delano2002pymol}
DeLano, W.~L. et~al.
\newblock Pymol: An open-source molecular graphics tool.
\newblock \emph{CCP4 Newsletter on protein crystallography}, 40\penalty0
  (1):\penalty0 82--92, 2002.

\bibitem[He et~al.(2016)He, Zhang, Ren, and Sun]{7780459}
He, K., Zhang, X., Ren, S., and Sun, J.
\newblock Deep residual learning for image recognition.
\newblock In \emph{2016 IEEE Conference on Computer Vision and Pattern
  Recognition (CVPR)}, pp.\  770--778, 2016.
\newblock \doi{10.1109/CVPR.2016.90}.

\bibitem[Jiménez et~al.(2018)Jiménez, Škalič, Martínez-Rosell, and
  De~Fabritiis]{doi:10.1021/acs.jcim.7b00650}
Jiménez, J., Škalič, M., Martínez-Rosell, G., and De~Fabritiis, G.
\newblock Kdeep: Protein–ligand absolute binding affinity prediction via
  3d-convolutional neural networks.
\newblock \emph{Journal of Chemical Information and Modeling}, 58\penalty0
  (2):\penalty0 287--296, 2018.
\newblock \doi{10.1021/acs.jcim.7b00650}.
\newblock URL \url{https://doi.org/10.1021/acs.jcim.7b00650}.
\newblock PMID: 29309725.

\bibitem[Koes et~al.(2013)Koes, Baumgartner, and Camacho]{koes2013lessons}
Koes, D.~R., Baumgartner, M.~P., and Camacho, C.~J.
\newblock Lessons learned in empirical scoring with smina from the csar 2011
  benchmarking exercise.
\newblock \emph{Journal of chemical information and modeling}, 53\penalty0
  (8):\penalty0 1893--1904, 2013.

\bibitem[Koshland~Jr(1995)]{koshland1995key}
Koshland~Jr, D.~E.
\newblock The key--lock theory and the induced fit theory.
\newblock \emph{Angewandte Chemie International Edition in English},
  33\penalty0 (23-24):\penalty0 2375--2378, 1995.

\bibitem[Lan et~al.(2020)Lan, Chen, Goodman, Gimpel, Sharma, and
  Soricut]{Lan2020ALBERT:}
Lan, Z., Chen, M., Goodman, S., Gimpel, K., Sharma, P., and Soricut, R.
\newblock Albert: A lite bert for self-supervised learning of language
  representations.
\newblock In \emph{International Conference on Learning Representations}, 2020.
\newblock URL \url{https://openreview.net/forum?id=H1eA7AEtvS}.

\bibitem[Le~Guilloux et~al.(2009)Le~Guilloux, Schmidtke, and
  Tuffery]{le2009fpocket}
Le~Guilloux, V., Schmidtke, P., and Tuffery, P.
\newblock Fpocket: an open source platform for ligand pocket detection.
\newblock \emph{BMC bioinformatics}, 10\penalty0 (1):\penalty0 1--11, 2009.

\bibitem[Son \& Kim(2021)Son and Kim]{10.1371/journal.pone.0249404}
Son, J. and Kim, D.
\newblock Development of a graph convolutional neural network model for
  efficient prediction of protein-ligand binding affinities.
\newblock \emph{PLOS ONE}, 16\penalty0 (4):\penalty0 1--13, 04 2021.
\newblock \doi{10.1371/journal.pone.0249404}.
\newblock URL \url{https://doi.org/10.1371/journal.pone.0249404}.

\bibitem[Stepniewska-Dziubinska et~al.(2018)Stepniewska-Dziubinska,
  Zielenkiewicz, and Siedlecki]{10.1093/bioinformatics/bty374}
Stepniewska-Dziubinska, M.~M., Zielenkiewicz, P., and Siedlecki, P.
\newblock {Development and evaluation of a deep learning model for
  protein–ligand binding affinity prediction}.
\newblock \emph{Bioinformatics}, 34\penalty0 (21):\penalty0 3666--3674, 05
  2018.
\newblock ISSN 1367-4803.
\newblock \doi{10.1093/bioinformatics/bty374}.
\newblock URL \url{https://doi.org/10.1093/bioinformatics/bty374}.

\bibitem[Trott \& Olson(2010)Trott and Olson]{Trott2010}
Trott, O. and Olson, A.~J.
\newblock Autodock vina: improving the speed and accuracy of docking with a new
  scoring function, efficient optimization, and multithreading.
\newblock \emph{Journal of computational chemistry}, 31\penalty0 (2):\penalty0
  455--461, Jan 2010.
\newblock ISSN 1096-987X.
\newblock \doi{10.1002/jcc.21334}.
\newblock URL \url{https://doi.org/10.1002/jcc.21334}.
\newblock PMID: 19499576 PMCID: PMC3041641.

\bibitem[Vamathevan et~al.(2019)Vamathevan, Clark, Czodrowski, Dunham, Ferran,
  Lee, Li, Madabhushi, Shah, Spitzer, and Zhao]{Vamathevan2019}
Vamathevan, J., Clark, D., Czodrowski, P., Dunham, I., Ferran, E., Lee, G., Li,
  B., Madabhushi, A., Shah, P., Spitzer, M., and Zhao, S.
\newblock Applications of machine learning in drug discovery and development.
\newblock \emph{Nature Reviews Drug Discovery}, 18\penalty0 (6):\penalty0
  463--477, Jun 2019.
\newblock ISSN 1474-1784.
\newblock \doi{10.1038/s41573-019-0024-5}.
\newblock URL \url{https://doi.org/10.1038/s41573-019-0024-5}.

\bibitem[Wang et~al.(2005)Wang, Fang, Lu, Yang, and
  Wang]{doi:10.1021/jm048957q}
Wang, R., Fang, X., Lu, Y., Yang, C.-Y., and Wang, S.
\newblock The pdbbind database: Methodologies and updates.
\newblock \emph{Journal of Medicinal Chemistry}, 48\penalty0 (12):\penalty0
  4111--4119, 2005.
\newblock \doi{10.1021/jm048957q}.
\newblock URL \url{https://doi.org/10.1021/jm048957q}.
\newblock PMID: 15943484.

\end{thebibliography}
\bibliographystyle{icml2021}

\end{document}